\documentclass[floatfix,prl,twocolumn,superscriptaddress,sn-aps,longbibliography]{revtex4-1}
\usepackage[utf8]{inputenc}

\usepackage{graphicx,epsfig,amsfonts}

\usepackage{bm}
\usepackage{amsmath}
\usepackage{amssymb}
\usepackage{float}
\usepackage{xcolor}
\usepackage[colorlinks=true,citecolor=blue]{hyperref} 
\usepackage{tabularx}
\usepackage{comment}

\DeclareGraphicsExtensions{.png,.jpg,.eps}

\usepackage{xcolor}

\begin{document}

\title{Hamiltonian learning of triplon excitations in an artificial nanoscale molecular quantum magnet}

\author{Rouven Koch}
\affiliation{QuTech and Kavli Institute of Nanoscience, Delft University of Technology, Delft 2628 CJ, The Netherlands}

\author{Robert Drost}
\affiliation{Department of Applied Physics, Aalto University, 02150 Espoo, Finland}

\author{Peter Liljeroth}
\affiliation{Department of Applied Physics, Aalto University, 02150 Espoo, Finland}

\author{Jose L. Lado}
\affiliation{Department of Applied Physics, Aalto University, 02150 Espoo, Finland}


\begin{abstract}
Extracting the Hamiltonian parameters of nanoscale quantum magnets from experimental measurements is a significant challenge in quantum matter. Here we establish a machine learning strategy to extract the parameters of a spin Hamiltonian from inelastic spectroscopy with scanning tunneling microscopy, and we demonstrate this methodology experimentally with an artificial nanoscale molecular magnet based on cobalt phthalocyanine (CoPC) molecules on NbSe$_2$. We show that this technique allows to extract the Hamiltonian parameters of a quantum magnet from the differential conductance, including the substrate-induced spatial variation of the exchange couplings. Our methodology leverages a machine learning algorithm trained on exact quantum many-body simulations with tensor networks of finite quantum magnets, leading to a methodology that predicts the Hamiltonian parameters of CoPC quantum magnets of arbitrary size. Our results demonstrate how quantum many-body methods and machine learning enable learning a microscopic description of nanoscale quantum many-body systems with scanning tunneling spectroscopy. 
\end{abstract}

\maketitle

  Quantum magnets represent one of the potential platforms to create exotic quantum excitations \cite{Sachdev2008,Vasiliev2018}. Quantum magnetism appears in Heisenberg models which are dominated by quantum fluctuations, an instance that often emerges in the presence of frustrated interactions \cite{Broholm2020,Kitaev2006,PhysRevLett.120.207203,PhysRevLett.123.207203,Chubukov1994,Yan2011}. These phenomena can give rise to a variety of excitations, including spinons, visons, gauge, and topological excitations \cite{Mourigal2013, DallaPiazza2014, Kohno2007,Mishra2021,Ruan2021,Zhao2024,Wang2024}. This should be contrasted with classical symmetry broken magnets featuring magnon excitations \cite{Spinelli2014,Klein2018,Cenker2020,PhysRevLett.123.047204,PhysRevX.9.011026,Ghazaryan2018,Ganguli2023,Chumak2015}. Understanding the nature of excitations of a specific quantum material, thereby telling quantum from classical magnets, requires knowledge of the underlying Hamiltonian, which is often exceptionally difficult to extract from experiments \cite{Takagi2019}.

\begin{figure}[t!]
\centering
\includegraphics[width=0.71\linewidth]{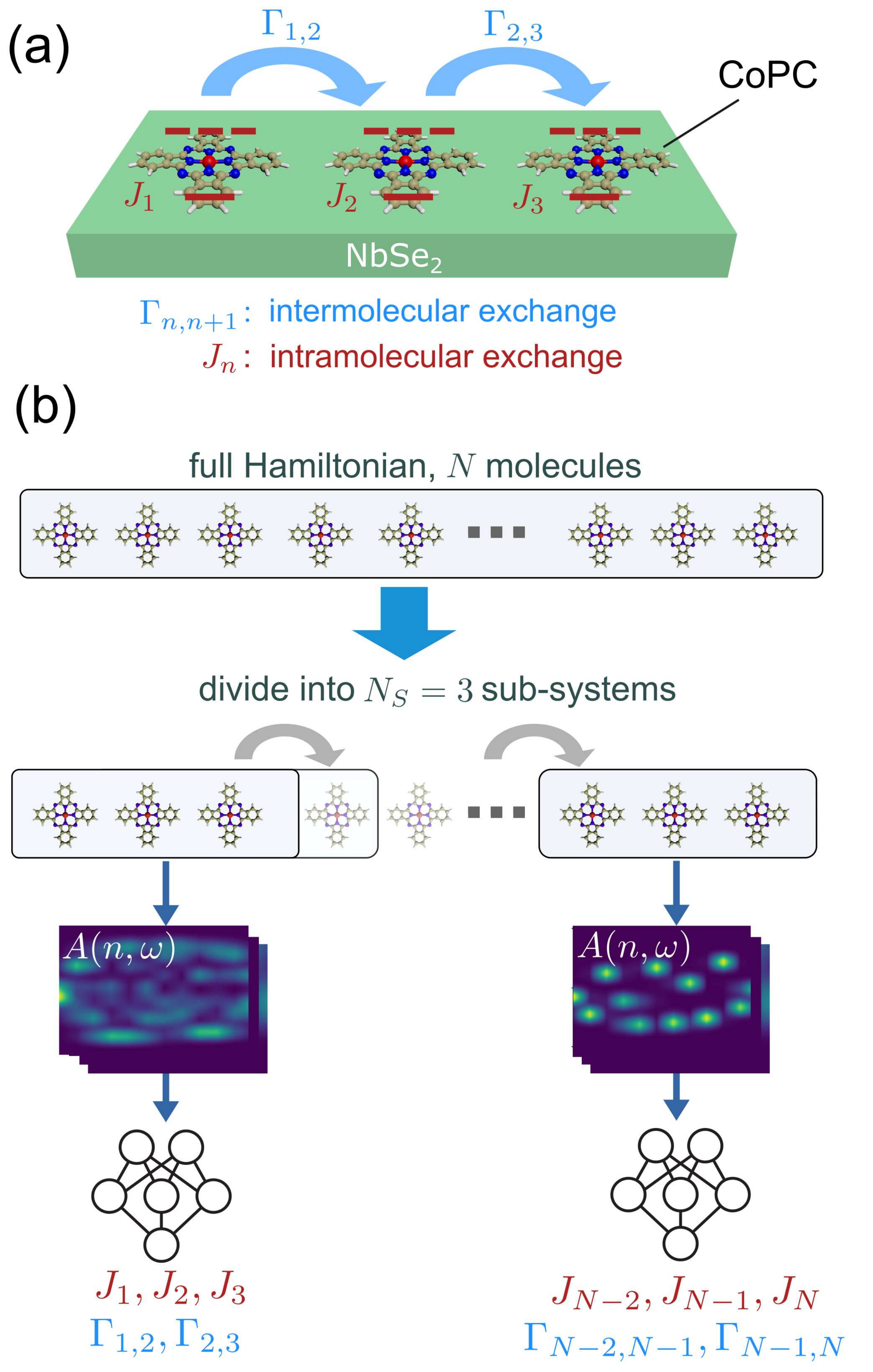}
\caption{
(a) Schematic of the one-dimensional spin model hosting triplon excitations with intra- and intermolecular exchange parameters $J$ and $\Gamma$. This system can be engineered in CoPC on NbSe$_2$. The red dashed lines represent the singlet and triplet states of the molecule.
(b) Machine learning workflow to extract Hamiltonian parameters of a chain of arbitrary length. Each site of the chain represents one CoPC molecule. The neural network predicts the spin Hamiltonian parameters of Eq.~\ref{eq:triplon}.
}
\label{fig: intro}
\end{figure}

Typical methodologies in quantum materials allow computing observables from Hamiltonians \cite{Bairey2019, Wang2017, Gebhart2023}.
However, extracting the Hamiltonian from a set of observables is often a challenging problem for conventional techniques. Machine learning provides a strategy to tackle such a complex inverse problem beyond the reach of conventional methodologies in quantum materials. This has been demonstrated for Hamiltonian learning with supervised learning \cite{Khosravian2024,2024arXiv241207666L,2024arXiv240504596V,Karjalainen2023} and generative machine learning \cite{PhysRevResearch.4.033223,Koch2023}, among others \cite{Gebhart2023,Valenti2022,Che2021,Garrison2018,Wang2017, simard2024}. However, learning Hamiltonians in quantum magnets remains a relatively unexplored problem, which ultimately may allow tackling the open challenge of identifying the nature of quantum spin liquids.

Here, we put forward a strategy to extract the underlying Hamiltonian parameters from scanning tunneling microscopy (STM) measurements of a molecular quantum magnet.
Our methodology relies on combining tensor-network many-body calculations of spin excitations of a molecular magnet with a machine learning methodology, which enables us to extract all Hamiltonian parameters of the system directly from local inelastic tunneling spectroscopy measurements.
The Hamiltonian learning algorithm can extract spatially dependent Hamiltonian parameters for arbitrarily large 1-dimensional molecular chains, solely by training the algorithm on many-body calculations of fixed-size systems. 
In particular, we demonstrate this methodology experimentally with a molecular quantum magnet hosting triplon excitations as realized in cobalt phthalocyanine (CoPC) molecules on NbSe$_2$. 
Our methodology puts forward a strategy to characterize with atomic resolution Hamiltonians of quantum magnets, including capturing local variations of exchange parameters, establishing a machine-learning-enabled technique for Hamiltonian learning in molecular quantum nanomagnetism.

\begin{figure}[t!]
\centering
\includegraphics[width=\linewidth]{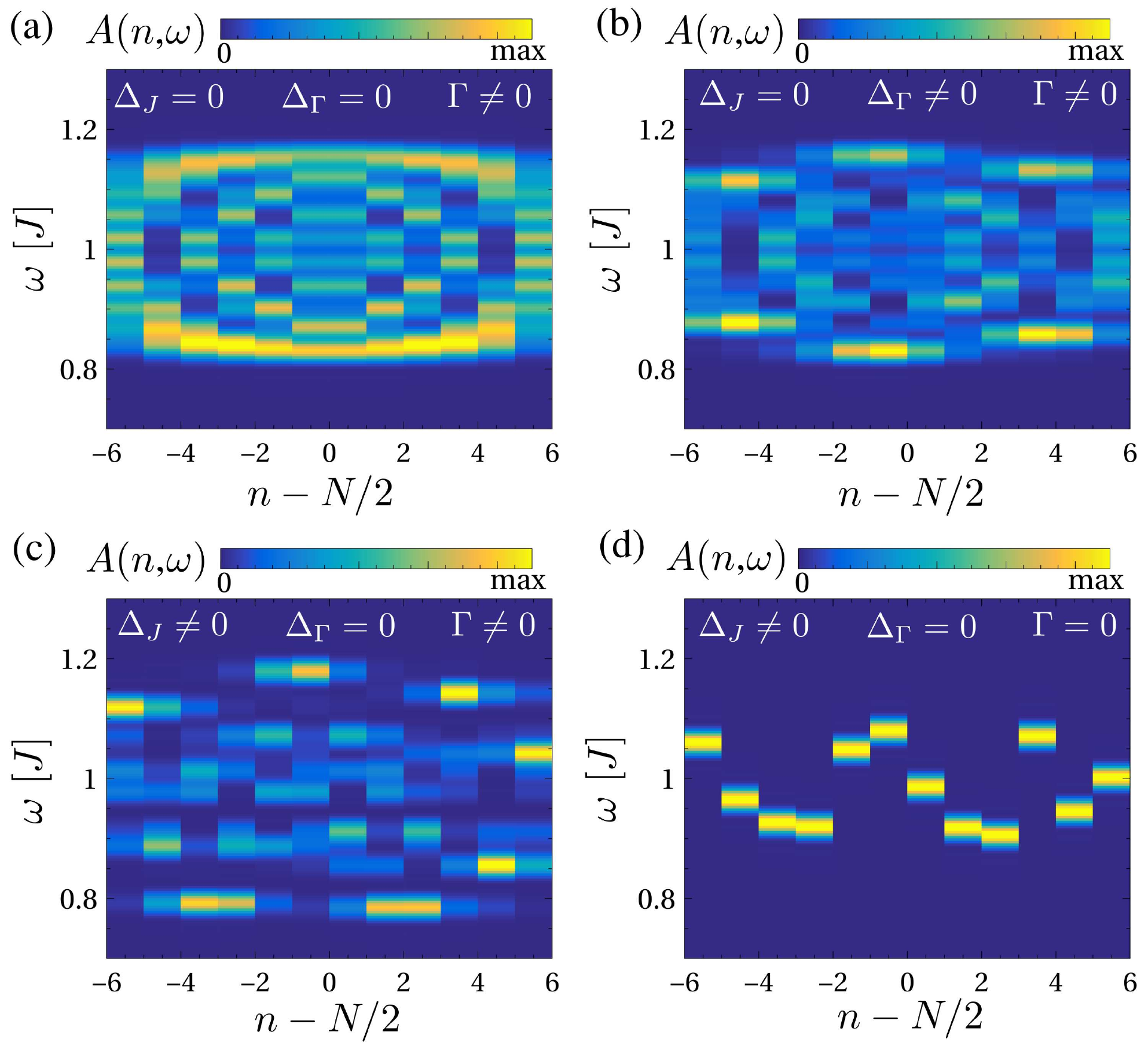}
\caption{Spectral function of the spin chain in the limit of no disorder
for the exchange couplings (a), finite coupling disorder (b), finite exchange disorder (c), and fully decoupled molecules (d).
It is observed that the presence of disorder and molecular coupling leads to a strongly featured spectral function.
We took $\Delta_J = 0.2J$, $\Delta_\Gamma = 0.2J$, and $\Gamma=0.3J$, and the same identical disorder profile was used in (b-d).
}
\label{fig:fig2}
\end{figure}

The system we focus on is a one-dimensional molecular quantum magnet that hosts triplon excitations and can be found experimentally in CoPC molecules on NbSe$_2$ (Fig.~\ref{fig: intro}(a)) \cite{Drost2023}. This system hosts two magnetic moments on two orbitals of the CoPC molecule, one in the center ion and the second distributed over the outer ligands~\cite{Wang2021}.
The molecular chain realizes the following Hamiltonian
\begin{equation}
\label{eq:triplon}
    H = \sum_n J_n \mathbf{S}_n \cdot \mathbf{K}_n + \sum_n \Gamma_{n,n+1} \mathbf{S}_n \cdot \mathbf{S}_{n+1}
\end{equation}
leading to a spin chain of length $N$ with a pair of spin-1/2 operators $\mathbf{K}_n = (\hat K^x_n,\hat K^y_n, \hat K_n^z)$ and $\mathbf{S} = (\hat S^x_n,\hat S^y_n, \hat S_n^z)$ on each molecule $n$.
The $J_n$'s are the intramolecular exchange couplings between the two orbitals and $\Gamma_{ij}$ is the intermolecular exchange coupling between neighboring molecules $n$ and $m$.
In general, the molecular chain will have average exchange couplings $J=\langle J_n\rangle$ and $\Gamma = \langle \Gamma_{n,n+1} \rangle$,
and random fluctuations of $\Delta_J$ and $\Delta_\Gamma$ around those averages. Triplons emerge in this system for $J\gg \Gamma$, where the bandwidth of the
triplon excitations scales as $\sim \Gamma$, and their gap as $\sim J$ \cite{Drost2023}.
The spectra of triplon excitations on site $n$ and frequency $\omega$ are accessed through the spectral function 
\begin{equation}
    A(n,\omega) = \sum_{\alpha=x,y,z}\langle GS | \hat{K}^\alpha_n \delta(\omega - \hat{H} + E_{GS}) \hat{K}^\alpha_n | GS \rangle
\end{equation}
of the many-body ground state $| GS \rangle$, that we compute using a tensor network kernel polynomial formalism \cite{dmrgpy,10.21468/SciPostPhysCodeb.4,PhysRevB.83.195115,PhysRevResearch.1.033009}. Inelastic spectroscopy on the molecule with STM measurements \cite{Spinelli2014,Heinrich2004,Kezilebieke2019,Toskovic2016,Mishra2021,RevModPhys.91.041001,PhysRevLett.122.227203} allows to directly access the previous spectral function as given by \cite{PhysRevLett.102.256802}
\begin{equation}
    A(n, \omega) \propto d^2I/dV^2 \, .
\end{equation} 
Typical spectra on the different sites of a molecular chain are shown in Fig.~\ref{fig:fig2}, where we show the limit of a pristine molecular chain $\Delta_J = 0$ (Fig.~\ref{fig:fig2}a), disorder in the coupling $\Delta_\Gamma\ne 0$ (Fig.~\ref{fig:fig2}b), disorder in the internal exchange $\Delta_J\ne 0$ (Fig.~\ref{fig:fig2}c), and a disordered decoupled chain with $\Gamma = 0$, $\Delta_J \ne 0$ (Fig.~\ref{fig:fig2}d). We observe that while both types of disorders $\Delta_J$ and $\Delta_\Gamma$ create fluctuations in the triplons, their relative values are challenging to directly extract from the spectral function.

\begin{figure}[t!]
\centering
\includegraphics[width=.99\linewidth]{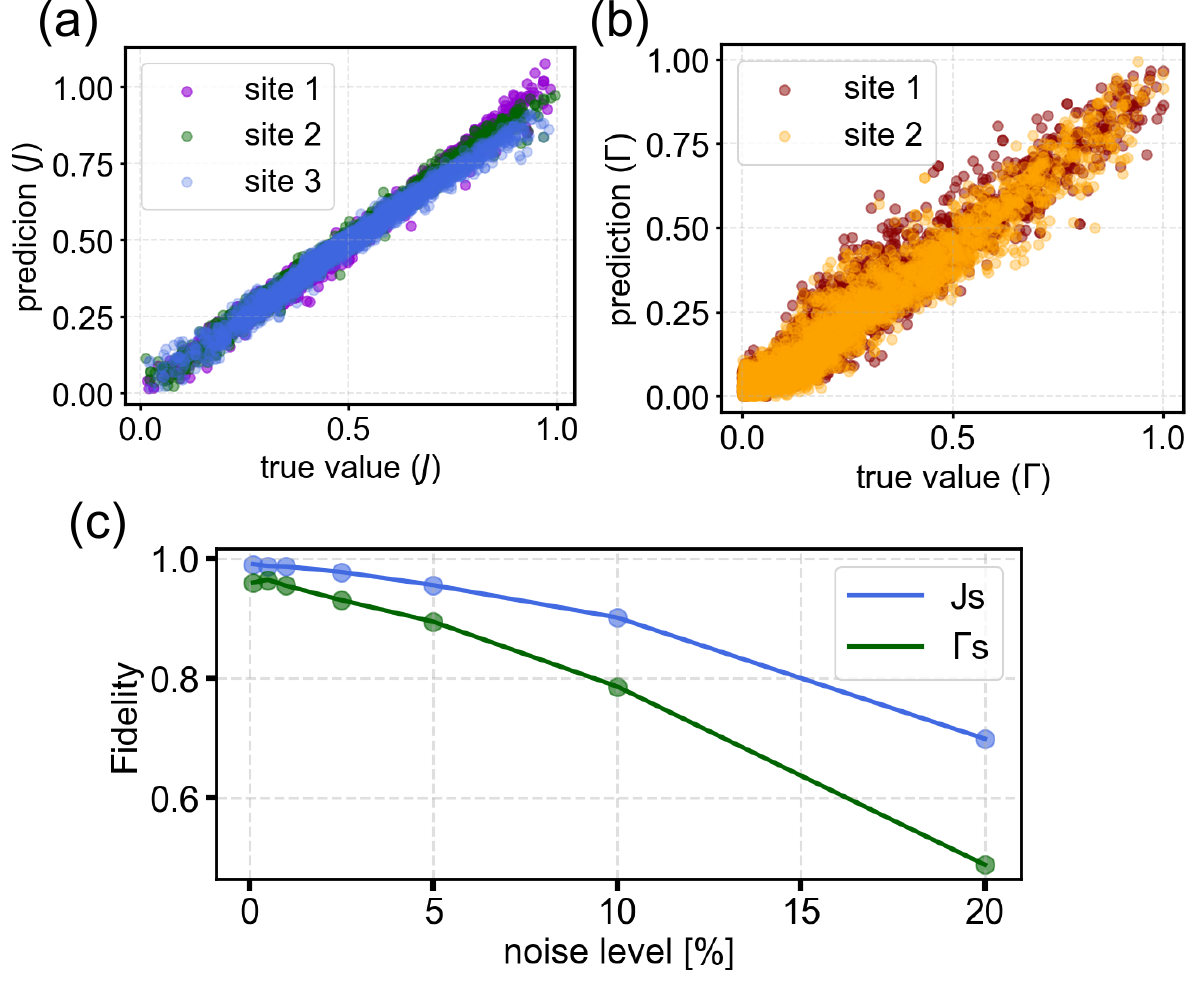}
\caption{
Panels (a) and (b) show the predictions of the NN algorithm to extract the intra- and intermolecular exchange amplitudes described in Eq.~\ref{eq:triplon} with the algorithm described in the Method section for 2$\%$ noise ($\eta=0.02$). The inputs of the model are $dI/dV$ spectra.
Panel (c) shows the fidelity (defined in Eq.~\ref{eq: fidelity}) vs. noise strength (defined in Eq.~\ref{noise}) for increasing noise.
}
\label{fig: MB_results}
\end{figure}

Our objective is to develop a machine learning algorithm that is trained on a finite system size that directly generalizes to smaller and larger systems to extract the Hamiltonian parameters of the molecular quantum magnet. The goal of our algorithm is to learn the underlying intra- and intermolecular exchange parameters, $J_n$ and $\Gamma_{n,n+1}$, by extracting information from the spectral functions that directly maps to $dI/dV$ measurements. For this, we develop an iterative workflow with a deep neural network (NN) as the central part to infer the  Hamiltonian parameters (details given in the SI).
We create a training set of molecular chains of length $N=12$ (24 spins $S=1/2$) with random Hamiltonian parameters $J_n$ and $\Gamma_{n,n+1}$. We choose $N=12$ to take a moderately large system where finite size effects are not dominating. Then, we separate the Hamiltonian of Eq.~\ref{eq:triplon} into sub-systems of length $N_S=3$ as depicted in Fig.~\ref{fig: intro}(b). The sub-system has five Hamiltonian parameters, three $J_n$ and two $\Gamma_{n,n+1}$. The NN algorithm predicts these five parameters at a time, taking the three $dI/dV$s (or spectral functions) of the sub-system molecules as inputs. Finally, the algorithm sweeps iteratively through the whole system of arbitrary size in slices of $N_S=3$ molecules. These slices, however, contain information from the larger parent Hamiltonian.
In addition, we average over the parameters of overlapping sites to obtain more accurate predictions. The workflow is illustrated in Fig.~\ref{fig: intro}(b). This enables us to make predictions for chains of arbitrary lengths by training the algorithm only on systems of size $N=12$ which keeps the computational costs very low even for predictions of very large systems. Our methodology allows us to directly extract spatially-dependent fluctuations of the exchange coupling, as emerging
from stacking dependent exchange in experimental molecular systems \cite{Drost2023}. 
In total, we create 1500 systems with varying disorder strength of size $N=12$ that we split up into $N_S=3$ subsystems to train the NN to infer the underlying parameters. Furthermore, we add noise to the simulated $dI/dV$ spectra in the form of
\begin{equation}\label{noise}
    \left(\frac{dI}{dV}\right)_{\text{noise}} = \left(\frac{dI}{dV}\right)_{\text{data}} + \eta \cdot \mathbf{R}
\end{equation}
  where $(dI/dV)_{\text{noise}}$ represents the noisy simulated differential conductance, $(dI/dV)_{\text{data}}$ is the original simulated differential conductance data (in form of an array with dimension [$\omega, N$]), \( \mathbf{R} \) is a random noise matrix of shape [$\omega, N$] uniformly distributed in \( [0, 1] \), \( \eta \) is the noise level
used in the interval \( [0, 0.2] \).
In Eq.~\ref{noise}, the $dI/dV$ is normalized to its maximum value, so that the normalized $dI/dV$ ranges between the minimum value 0 and its maximum value 1. Thus, the random term containing $\eta$ in Eq.~\ref{noise} controls the noise level as a percentage, where a value of $\eta=0.2$ corresponds to 20\% noise.  More details about the data modeling, creation of the dataset, and the post-processing of the experimental data can be found in the Supplementary Information (SI).
The ML model that we use is a feed forward NN, trained with data from the many-body spin chain from Eq.~\ref{eq:triplon}, where we define the intra-molecular exchange $J_n \in [0,1]$ and inter-molecular exchange $\Gamma_{n,n+1} \in [0, 0.4]$. 
The ML algorithm with the underlying NN learns to predict the underlying Hamiltonian of a $N_S = 3$-molecule sub-Hamiltonian by taking three spectral functions or $dI/dV$ spectra as inputs, as depicted in Fig.~\ref{fig: intro}(b). 
While being trained on sub-systems of the $N= 12$-molecule chain, the algorithm transfers to smaller and larger system sizes without decreased precision.
The details of the algorithm, including the architecture and training parameters, can be found in the SI and in Ref.~\cite{rouvencode}.

\subsection{Triplon excitations in a quantum magnet}

\begin{figure}[t!]
\centering
\includegraphics[width=1.0\linewidth]{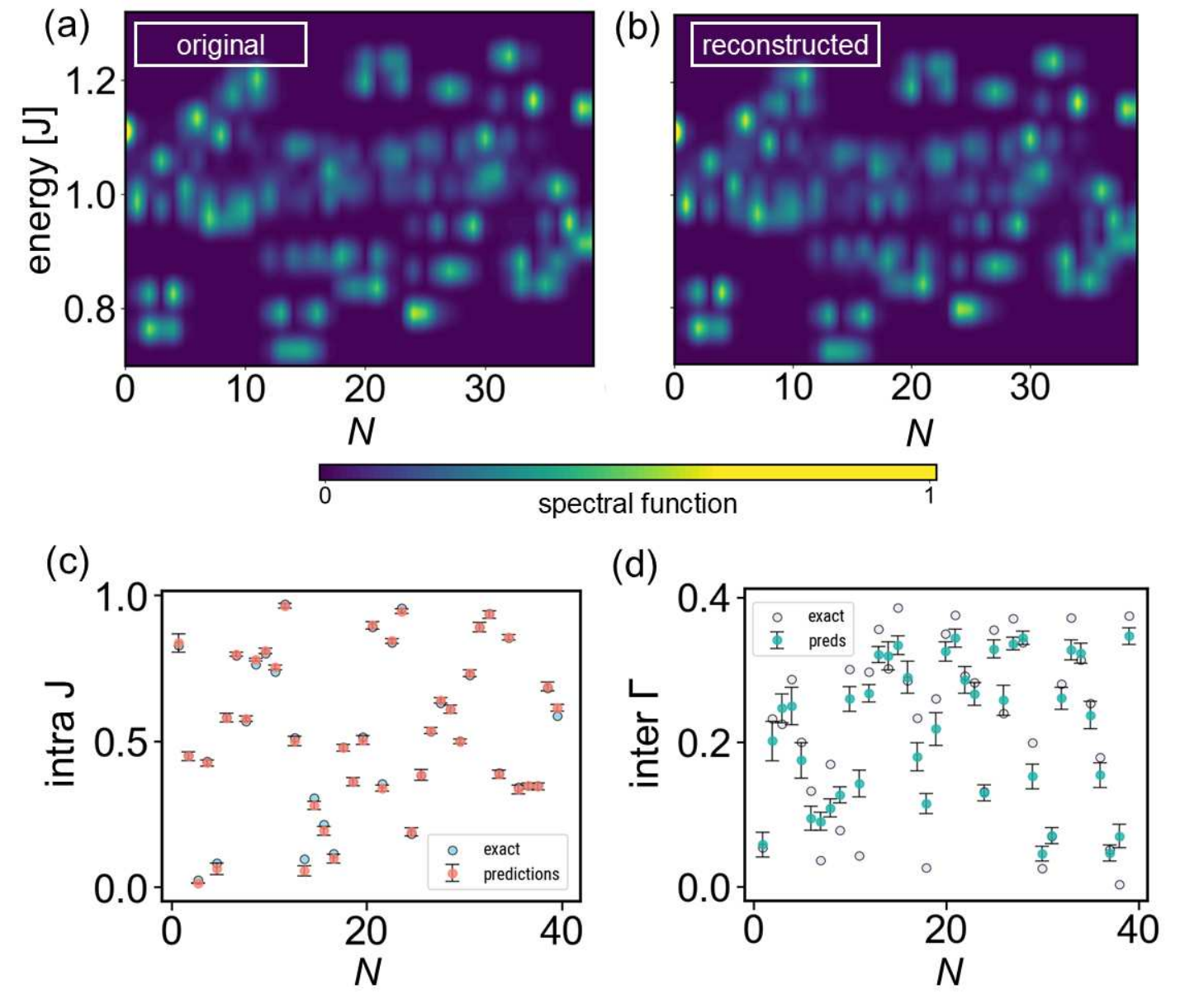}
\caption{Hamiltonian learning algorithm trained on systems of size $N=12$ applied to a simulated molecular chain of size $N=40$. Panels (a) and (b) show the original and reconstructed spectral function. Panels (c) and (d) show the extracted and exact Hamiltonian parameters for the intra- and intermolecular exchange. The predictions are averaged over 10 random initializations of the NN.}
\label{fig: n40_results}
\end{figure}

In Fig.~\ref{fig: MB_results}, we demonstrate the performance of the ML algorithm on the test data of the $N=12$ many-body model (for added noise of 2$\%$).
We compare the predictions of the intra-molecular $J_n$ (a) and inter-molecular $\Gamma_{n,n+1}$ exchange in Fig.~\ref{fig: MB_results}(b) with their true value. 
The test samples are divided into sub-Hamiltonians of size $N_S=3$.
We observe that the algorithm predicts the intra-molecular exchange $J_n$ in Fig.~\ref{fig: MB_results}(a) with high accuracy, showing small deviations from the ideal match and a mean absolute error (MAE) of $\mathcal{E}_J=0.024$.
The predictions of the inter-molecular exchange $\Gamma_{n,n+1}$ Fig.~\ref{fig: MB_results}(b) show similar behavior, with slightly higher deviations from the ideal match and an increased MAE of $\mathcal{E}_\Gamma=0.051 $.
The intramolecular exchange $J_n$ determines the position of the excitation spectra and the intermolecular exchange $\Gamma_{n,n+1}$ determines the width.
The difference in accuracy of the predictions is related to the higher complexity and impact of $\Gamma_{n,n+1}$ on the shape and features of the molecular chain.

The quality of the Hamiltonian extraction can be characterized by the fidelity between the prediction and the real exchange couplings defined
as \cite{Khosravian2024}

\begin{equation} \label{eq: fidelity}
\begin{split}
& F(\Lambda_{\text{pred}}, \Lambda_{\text{true}}) = \\[1mm]
& = \frac{\left| \langle \Lambda_{\text{pred}} \Lambda_{\text{true}} \rangle - \langle \Lambda_{\text{pred}} \rangle \langle \Lambda_{\text{true}} \rangle \right|}
{\sqrt{\left( \langle \Lambda_{\text{true}}^2 \rangle - \langle \Lambda_{\text{true}} \rangle^2 \right) 
\left( \langle \Lambda_{\text{pred}}^2 \rangle - \langle \Lambda_{\text{pred}} \rangle^2 \right)}}
\end{split}
\end{equation}
where $\Lambda_{\text{true}}= J_n^{\text{true}},\Gamma_{n,n+1}^{\text{true}}$ and $\Lambda_{\text{pred}}= J_n^{\text{pred}},\Gamma_{n,n+1}^{\text{pred}}$ are the true and predicted Hamiltonian parameters of the system. 
{We calculate the fidelity on the simulated test data and, therefore, include the ensemble average ⟨x⟩ over the whole test set.} 
The fidelity is defined in the interval $\mathcal{F} \in [0,1]$ where 1 stands for identical predictions and true values and 0 for fully uncorrelated values.
In Fig.~\ref{fig: MB_results}(c), we show the resilience of the algorithm to noise added to the data. 
Fig.~\ref{fig: MB_results}(c) shows that even for high noise levels of more than $10\%$, the fidelity of the intra- and intermolecular exchange remains high. As expected from the results of Fig.~\ref{fig: MB_results}(a,b), the predictions for the intramolecular exchange have generally higher fidelity.

Now, we demonstrate that our algorithm is capable of extending to significantly longer spin chains.
We apply the algorithm that is trained on systems of size $N=12$ to a simulated $N=40$ molecular spin chain with randomly chosen $J_n$ and $\Gamma_{n,n+1}$ and predict the underlying parameters, divided into $N_S=3$-molecule sub-systems.
In Fig.~\ref{fig: n40_results}(a,b), we compare the calculated spectral function with the reconstructed one and show the difference in predictions for the intra- and intermolecular exchange in Fig.~\ref{fig: n40_results}(c,d).
We find that we can extract the intramolecular exchange with very high precision and a significantly lower error than the intramolecular exchange. 
These results are in accordance with the findings for the $N=12$ molecule systems of Fig.~\ref{fig: MB_results}(a,b) where we discuss that $\Gamma_{n,n+1}$ has a significantly lower impact on the spectral function and $dI/dV$ compared to $J_n$ and, therefore, is inherently more difficult to determine regardless of the method. 
However, the appearance of the reconstructed molecular chain Fig.~\ref{fig: n40_results}(b) is almost indistinguishable from the original calculation Fig.~\ref{fig: n40_results}(a) that is used as input to infer the parameters. This highlights that $J$ has the greatest impact on the main features of the spectral functions (and $dI/dV$).
These results demonstrate that our algorithm is capable of extending to significantly longer chains and gives faithful results for arbitrarily long chains.

\begin{figure}[t!]
\centering
\includegraphics[width=.99\linewidth]{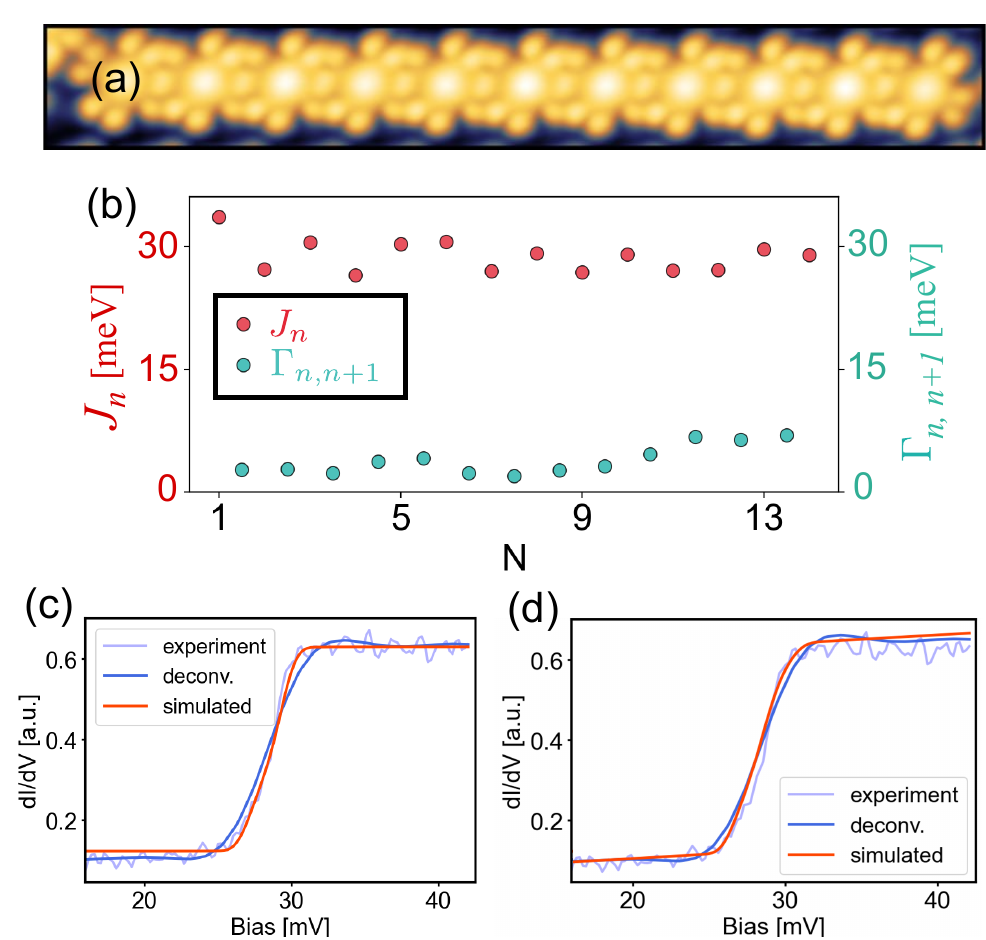}
\caption{(a) Image of a CoPC molecular chain on NbSe$_2$ (size $20\times3$nm$^2$).
(b) Extracted $J_n$ and $\Gamma_{n,n+1}$ for the $N=14$ molecular chain. 
(c,d) parameter extraction and reconstructed $dI/dV$ from STM measurements. The examples are taken from $dI/dV$ spectra from the chains with $N=14$ (c) and $N=11$ molecules (d). The experimental and deconvolved spectra are compared to the reconstructed simulated $dI/dV$. 
}
\label{fig: HL experiments}
\end{figure}

\subsection{Application of the algorithm to experimental molecular chain}
We now apply the algorithm to real measurement data for a one-dimensional molecular chain. The sample was fabricated by subliming CoPc molecules onto a freshly cleaved NbSe$_2$ substrate. Subsequently, the sample was transferred to a low-temperature STM operating at 4~K. The measurements were performed with a NbSe$_2$-coated superconducting tip to enhance the energy resolution \cite{PhysRevLett.100.226801,Franke2011,Kezilebieke2018}. This induces sharp peaks at the edges of the spin-flip excitations. The spectra can be deconvolved with the tip spectral density to remove this effect (details are given in the SI). Depending on the surface coverage, CoPC self-assembles into various motifs on NbSe$_2$, forming individual molecules, molecular chains, and islands \cite{Drost2023}. 

We now use our machine learning methodology to the experimental
molecular chain. A typical STM image of such a system is shown in Fig.~\ref{fig: HL experiments}(a). 
In Fig.~\ref{fig: HL experiments}(b), we show the results of the Hamiltonian learning algorithm applied to the $N=14$ molecular chain, 
specifically enabling the extraction of the intra- and intermolecular exchange couplings for each molecule on the chain. 
In Fig.~\ref{fig: HL experiments}(c,d), we show example spectra from the measured triplon chains of lengths $N=14$ and $N=11$, respectively. 
The Hamiltonian parameters are extracted from the deconvolved $dI/dV$ spectra. We compare the reconstructed $dI/dV$ spectra with the experimental and deconvolved spectra~\footnote{Offset and absolute magnitude of the simulated data are adjusted to emulate the experimental spectra, as outlined in the SI}. 
The results demonstrate that the weight of the step, which is proportional to the exchange coupling constant ($J_n$), is accurately captured in panels Fig.~\ref{fig: HL experiments}(c,d). Furthermore, the width and steps, which are related to the broadening parameter ($\Gamma_{n,n+1}$), are also well reproduced for both chains ($N=14,11$)~\footnote{As seen in Fig.~\ref{eq: fidelity}(b), the predictions for the intermolecular exchange show a higher deviation. This is also the case for the experimental $dI/dV$ spectra, i.e., the center of the step ($J_n$) is predicted with high precision and the width and steps ($\Gamma_{n,n+1}$) can deviate more, depending on the chain and specific measurement.}.
For the intermolecular exchange ($\Gamma_{n,n+1}$), we obtain values of $\Gamma \sim 0.08-0.19 J$, depending on the system size and specific molecule, consistent with previous average estimates \cite{Drost2023}. 
These findings highlight that training the ML model with simulated data generalizes effectively to experimental data, eliminating the need for re-training, and is capable of predicting Hamiltonian parameters in systems whose system size is different from the theoretical training set. 
As a result, our algorithm is system-size independent and can be applied to experimental systems of arbitrary size, provided $N > 3$.

To summarize, here we presented a machine learning strategy to extract the underlying Hamiltonian of one-dimensional molecular spin systems. This methodology was demonstrated using a molecular spin chain hosting triplon excitations, highlighting its ability to extract information of 1D chains of arbitrary length by dividing the system into sub-Hamiltonians. 
Our methodology performs a faithful Hamiltonian extraction across a wide range of systems, from those with no disorder to highly disordered configurations.
We showed that by solely training our algorithm in chains with $N=12$ molecules, the machine learning method enables us to perform Hamiltonian learning in systems of arbitrary size, in particular with $N=40$ emulated molecular spin chains.
We applied our strategy for Hamiltonian learning to experimental $dI/dV$ measurements of molecular quantum magnets, where we show accurate results in extracting Hamiltonian parameters in disordered systems in the presence of noise in chains up to $N=14$ molecules.
This strategy allows us to train the algorithm to work with arbitrary system sizes using quantum many-body calculations of specific finite-size systems.
This approach can be extended to general spin models beyond those featuring triplons and general one-dimensional many-body Hamiltonians, possibly even to more spatial dimensions.
While extending to two dimensions is more challenging due to the rapid growth of entanglement entropy~\cite{PhysRevLett.109.067201,Zheng2017,Ors2019}, emerging numerical techniques such as Neural Quantum States offer promising alternatives for generating training data in two-dimensional systems~\cite{Carleo2017,PhysRevB.100.125124,Hermann2023}.
Our results establish a versatile framework to perform Hamiltonian learning in engineered molecular quantum magnets, which can be extended to generic quantum lattice models, including interacting quantum dots or qubit arrays.\\

  \textbf{Supporting Information:} \\
Details of data modeling of the simulations, post-processing of experimental data, experimental and measurement setup, neural network architecture, training parameters, and inference, and analysis of broadening effects on the Hamiltonian extraction \\ 

  \textbf{Author Information:} \\
Rouven Koch - QuTech and Kavli Institute of Nanoscience, Delft University of Technology, Delft 2628 CJ, The Netherlands; orcid.org/0000-0002-8917-3631; Email: r.k.koch@tudelft.nl\\[1mm]
Robert Drost - Department of Applied Physics, Aalto University,
02150 Espoo, Finland; Email: robert.drost@aalto.fi\\[1mm]
Peter Liljeroth - Department of Applied Physics, Aalto University,
02150 Espoo, Finland; Email: peter.liljeroth@aalto.fi\\[1mm]
Jose L. Lado - Department of Applied Physics, Aalto University,
02150 Espoo, Finland; Email: jose.lado@aalto.fi\\

  \textbf{Acknowledgments:}\\
This research made use of the Aalto Nanomicroscopy Center (Aalto NMC) facilities and was supported by the Academy of Finland Projects Nos. 331342, 358088, 368478, 353839, and 347266, the Finnish Quantum Flagship, ERC AdG GETREAL (no. 101142364),
ERC CoG ULTRATWISTROICS (no. 101170477), and the KIND synergy program from the Kavli Institute of Nanoscience Delft. We thank S. Kezilebieke for help during the early stages of this project. We acknowledge the computational resources provided by the Aalto Science-IT project.

\section*{References}

\bibliography{references}

 \end{document}